\documentclass[12pt]{article}
\usepackage{color}

\newcommand{\beq}{\begin{equation}}
\newcommand{\eeq}{\end{equation}}
\newcommand{\beqa}{\begin{eqnarray}}
\newcommand{\eeqa}{\end{eqnarray}}
\newcommand{\beqar}{\begin{eqnarray*}}
\newcommand{\eeqar}{\end{eqnarray*}}

\newcommand{\al}{\alpha}
\newcommand{\be}{\beta}

\def\spa          {\ \ \ }
\def\non          {\nonumber}
\def\ha           {\mbox{$\frac{1}{2}$}}

\def\s  {\sigma}
\def\spa          {\ \ \ }
\def\mand         {\spa\mbox{and}\spa}

\def\Tr           {\mbox{\rm Tr}\,}
\def\STr          {\mbox{\rm STr}\,}

\def\cd           {{\cdot}}
\def\ran          {\rangle}
\def\lan          {\langle}
\def\fsH    {H\!\!\!\!/\,}
\newcommand{\del}{\delta}

\newcommand{\eps}{\epsilon}
\newcommand{\ga}{\gamma}

\newcommand{\inn}{\!\cdot\!}

\newcommand{\lam}{\lambda}

\newcommand{\z}{\zeta}

\newcommand{\ie}{{\it i.e.,}\ }
\newcommand{\labell}[1]{\label{#1}} 
\newcommand{\reef}[1]{(\ref{#1})}
\newcommand\prt{\partial}
\newcommand\veps{\varepsilon}

\newcommand\cL{{\cal L}}
\newcommand\cD{{\cal D}}

\newcommand\bz{\bar{z}}

\def\sst#1{{\scriptscriptstyle #1}}
\def\0{{\sst{(0)}}}
\def\1{{\sst{(1)}}}
\def\2{{\sst{(2)}}}
\def\3{{\sst{(3)}}}
\def\4{{\sst{(4)}}}
\def\5{{\sst{(5)}}}
\def\6{{\sst{(6)}}}
\def\7{{\sst{(7)}}}
\def\8{{\sst{(8)}}}

\begin{document}
\baselineskip 18pt%
\begin{titlepage}
\vspace*{1mm}%
\hfill
\vbox{

    \halign{#\hfil         \cr
         ICTP-PH-TH/2012-xyz\cr
           } 
      }  
\vspace*{8mm}
\vspace*{8mm}%

\center{ {\bf \Large More on closed string induced higher derivative
interactions on D-branes

}}\vspace*{3mm} \centerline{{\Large {\bf  }}}
\vspace*{5mm}
\begin{center}
{Ehsan Hatefi$^{\,\dagger}$ and I. Y. Park$\,^*$}

\vspace*{0.8cm}{ {\it
International Centre for Theoretical Physics\\
 Strada Costiera 11, Trieste, Italy $^{\dagger}$ \\
ehatefi@ictp.it}}

\vspace*{0.4cm}
{\it Department of Natural and physical Sciences,
Philander Smith College
                               \\
Little Rock, AR 72223, USA $^*$ \\
inyongpark05@gmail.com
}

\vspace*{.3cm}
\end{center}
\begin{center}{\bf Abstract}\end{center}
\begin{quote}
In our continued efforts of matching full string computations
with the corresponding effective field theory computations, we evaluate
 string theory correlators in closed forms. In particular, we consider
a correlator between
 three SYM vertex operators and one Ramond-Ramond $C$-field
vertex operator:
\vskip 0.001in
$<V_{C}V_{\phi} V_AV_A>$.
We show
that the infinite number of massless poles
of this amplitude can be reproduced by the Born-Infeld action, the Wess-Zumino terms, and their higher derivative corrections.
More specifically we find, up to an on-shell ambiguity, two scalar field  and two gauge field couplings to all orders in $\alpha'$ such that the infinite number of massless poles of the field theory amplitude exactly match the infinite number of massless poles of S-matrix elements of $<V_{C}V_{\phi} V_AV_A>$. We comment on close intertwinedness of an open string and
a closed string that must be behind the matching.
\end{quote}
\end{titlepage}

\section{Introduction}

D-brane physics \cite{Polchinski:1995mt}\cite{Witten:1995im}\cite{Polchinski:1996na}
 has played a central role in theoretical high energy
physics for more than one and a half decades by now. In the open string
description, a D$_p$-brane with a $(p+1)$-dimensional world volume
is realized as a hypersurface in flat
spacetime with the appropriate boundary condition on the string coordinates:
Dirichlet boundary conditions on the directions transverse
 to the D$_p$-brane and Neumann boundary conditions along the worldvolume of the D$_p$-brane \cite{Polchinski:1994fq}.
The bosonic action for multiple D$_p$-branes  was given by Myers \cite{Myers:1999ps}.
Note that a supersymmetric generalization is still unknown; see however \cite{Howe:2006rv}. The effective action for a single bosonic D$_p$-brane  was found in \cite{Leigh:1989jq}. The supersymmetric action for a single D$_p$-brane  was derived in \cite{Cederwall:1996pv}. See \cite{Hatefi:2010ik} for more details on Born-Infeld, Chern-Simons actions and their higher derivative corrections.
 Section 5 of  \cite{Hatefi:2012wj} has a review of Chern-Simons action.

\vskip 0.1in
 The advent of D$_p$-brane physics
has greatly promoted the significance of an open
 string bringing numerous new results, and is likely to
continue
bringing exciting new physics in the future.
 A conjecture put forward in \cite{Park:2007mc} may be an
example: quantum
effects of open strings moving on
D$_p$-branes should produce the curvature of the host
D$_p$-branes.

\vskip 0.1in

Having concrete tools for various multi-point string amplitudes
is important for many purposes, including the first-principle
derivation of AdS/CFT.
Previous works on scattering that involve D$_p$-branes
and some applications of D$_p$-branes include \cite{Hashimoto:1996bf}.
Given that a close interplay between an
open string and a closed string must be behind AdS/CFT, amplitudes
involving a mixture of open string states and closed string states
should be especially worth studying.
In this work, we continue our previous
endeavours of computing amplitudes of one Ramond-Ramond
$C$-field vertex
and three massless open string vertices \cite{Hatefi:2010ik}.
 The amplitudes that we specifically consider are
$<V_{C}V_{\phi} V_A>$ and $<V_{C}V_{\phi} V_AV_A>$.\footnote{
One may wonder whether it would be possible by applying string
T-duality to deduce
the result of $<V_{C}V_{\phi} V_AV_A>$ from \cite{Hatefi:2010ik} in which
$<V_{C}V_A V_AV_A>$ was analysed. Applying T-duality in present
 types of computations can be subtle. It is definitely so in loop-level
computations \cite{Park:2008sg}. Although the present computations are all at tree-level
the presence of a closed string state can make things subtle, and
indeed does so.
}
(We focus on  $<V_{C}V_{\phi} V_AV_A>$ in the main text presenting
the simpler case $<V_{C}V_{\phi} V_A>$ in Appendix B.)

\vskip 0.1in

Applying the same methodology as that of the previous works \cite{Hatefi:2010ik},\cite{Garousi:2007fk} and \cite{Hatefi:2008ab},
we find below the precise (up to an on-shell ambiguity) forms
of the two scalar two gauge field vertices such that the infinite number of massless poles of the field theory amplitude exactly match the infinite number of massless poles of S-matrix elements of $<V_{C}V_{\phi} V_AV_A>$ to all
orders in $\alpha'$.
 To obtain the infinite number of massless
  poles for this case, one needs to determine the higher derivative couplings of  two scalar two gauge field couplings to all
orders in $\alpha'$. (This is analogous to the situation in \cite{Hatefi:2010ik} where
$<V_{C}V_{A} V_AV_A>$ was analyzed. The precise form of the four gauge field
vertices was determined for the field theory amplitudes to match with the corresponding string theory computations.)
We obtain all the infinite number of massless poles in field theory
 and find consistent results with the string theory counterpart.

\vskip 0.1in

Additionally, remarks on a subtlety are in order. The subtlety may
be tied with the profound relation
between an open and closed string.
The present work is another demonstration that the pure SYM vertices (such as two scalar two gauge field couplings of this work) produce
the same massless poles as the corresponding
correlator of one RR  field vertex $V_C$ and three SYM vertices.

\vskip 0.1in

This phenomenon seems quite universal and
must have deep origins going back to the intertwinedness
of an open string and a closed string. The intertwinedness should
originate from the composite nature of a closed string state in terms of open string
states. The compositeness is (implicitly) exploited in a common
practice in literature, identification of both sets of closed oscillators
with those of an open string's.
  We will briefly comment on this issue in section 2 and contemplate
the issue further in the conclusion.

\vspace{.3in}
The organization of the paper is as follows. In section 2, we calculate a tree-level four-point string scattering of one RR vertex, one
scalar field vertex operator and two gauge field vertex operators, $<V_{C}V_{\phi} V_AV_A>$.
 In section 3, we consider
the low energy effective field theory and determine
the interaction vertices that, with Myers' terms,
produce the same massless poles as those of the string amplitude.
In the conclusion, we summarize the results and ponder on the profound relation between
an open string and a closed string as a reason for the matching. We end with comments on the future
directions.
In Appendix A, a summary of our conventions is presented.
Appendix B contains a parallel analysis for the case of
$<V_{C}V_{\phi} V_A>$.

\section{String amplitude computations }

By applying the conformal field theory technique, we carry out the string scattering amplitude of one closed string Ramond-Ramond  field,
one scalar field  and two gauge fields on the world volume of
BPS D$_p$-branes in type II super string theory within  a flat background.
Some efforts for the tree level scattering amplitudes have been done
\cite{Stieberger:2009hq,Kennedy:1999nn,Chandia:2003sh,Hatefi:2010ik,Garousi:2007fk,Hatefi:2008ab}.
To compute a S-matrix element, one has to know  the picture of the vertex operators so that the sum of the super ghost charges must be -2 for disk level amplitudes.

The vertex operators that we will need are given by\footnote{We keep $\alpha'$
explicitly in this work. One may set $\alpha'=2$ to simplify the expressions.}
\beqa
V_{\phi}^{(0)}(x) &=& \xi_{i}\bigg(\partial
X^i(x)+\alpha' ik\cd\psi\psi^i(x)\bigg)e^{\alpha' ik\cd X(x)},
\nonumber\\
V_{\phi}^{(-1)}(y) &=&\xi.\psi(y) e^{-\phi(y)} e^{\alpha' ik\cd X(y)},
\nonumber\\
V_{A}^{(0)}(x) &=& \xi_{a}\bigg(\partial
X^a(x)+ \alpha'iq\cd\psi\psi^a(x)\bigg)e^{ \alpha' iq\cd X(x)},
\nonumber\\
V_{A}^{(-1)}(y) &=&\xi_a\psi^a(y) e^{-\phi(y)} e^{\alpha'iq\cd X(y)}
\nonumber\\
V_{C}^{(-\frac{1}{2},-\frac{1}{2})}(z,\bar{z})&=&(P_{-}\fsH_{(n)}M_p)^{\al\be}e^{-\phi(z)/2}
S_{\al}(z)e^{i\frac{\alpha'}{2}p\cd X(z)}e^{-\phi(\bar{z})/2} S_{\be}(\bar{z})
e^{i\frac{\alpha'}{2}p\cd D \cd X(\bar{z})},
\label{d4Vs}
\eeqa
where $(k,q,p)$ are the momenta
of the scalar field, gauge field and $C$-field respectively; they satisfy the on-shell condition
$k^2=q^2=p^2=0$.
Our notation is such that
 the spinorial
indices are raised by the charge conjugation matrix, $C^{\alpha\be}$
\beqa
(P_{-}\fsH_{(n)})^{\al\be} =
C^{\al\del}(P_{-}\fsH_{(n)})_{\del}{}^{\be}
\eeqa
In particular, the trace is defined by
\beqa
\Tr(P_{-}\fsH_{(n)}M_p\gamma^{k})&\equiv & (P_{-}\fsH_{(n)}M_p)^{\alpha\beta}(\gamma^{k}C^{-1})_{\alpha\beta}
\nonumber\\
\Tr(P_{-}\fsH_{(n)}M_p\Gamma^{jai})&\equiv &
(P_{-}\fsH_{(n)}M_p)^{\alpha\beta}(\Gamma^{jai}C^{-1})_{\alpha\beta}
\eeqa
where $P_{-}$ is a projection operator, $P_{-} = \ha (1-\ga^{11})$, and
\begin{displaymath}
\fsH_{(n)} = \frac{a
_n}{n!}H_{\mu_{1}\ldots\mu_{n}}\ga^{\mu_{1}}\ldots
\ga^{\mu_{n}}
\ ,
\non\end{displaymath}
with $n=2,4$ for type IIA and $n=1,3,5$ for type IIB. $a_n=i$ for IIA and $a_n=1$ for IIB theory.
To employ standard holomorphic worldsheet correlators,
we implement the usual doubling trick. For more details see Appendix A of \cite{Hatefi:2012wj}.

\vskip 0.2in

\subsection{Computation of  $<V_{C}V_{\phi} V_A V_A>$ }

The S-matrix element of one closed string C field, one scalar field and two gauge fields
is given by the following correlation function

\begin{eqnarray}
{\cal A}^{C \phi AA} & \sim & \int dx_{1}dx_{2}dx_{3}dzd\bar{z}\,
  \lan V_{\phi}^{(-1)}{(x_{1})}
V_{A}^{(0)}{(x_{2})}V_A^{(0)}{(x_{3})}
V_{RR}^{(-\frac{1}{2},-\frac{1}{2})}(z,\bar{z})\ran,\labell{sstring}\eeqa
The open string vertex operators are inserted
at the boundary of the disk worldsheet while the closed string
vertex operator is inserted inside.
The  amplitude  reduces to the following
correlators for 123 ordering
\beqa {\cal A}^{C \phi AA}&\sim& \int
 dx_{1}dx_{2}dx_{3}dx_{4} dx_{5}\,
(P_{-}\fsH_{(n)}M_p)^{\al\be}\xi_{1i}\xi_{2a}\xi_{3b}x_{45}^{-1/4}(x_{14}x_{15})^{-1/2}\nonumber\\&&
\times(I_1+I_2+I_3+I_4)\Tr(\lam_1\lam_2\lam_3),\labell{125}\eeqa where
$x_{ij}=x_i-x_j$, and
\beqa
I_1&=&{<:e^{\alpha' ik_1.X(x_1)}:\partial X^a(x_2)e^{\alpha' ik_2.X(x_2)}
:\partial X^b(x_3)e^{\alpha' ik_3.X(x_3)}:e^{i\frac{\alpha'}{2}p.X(x_4)}:e^{i\frac{\alpha'}{2}p.D.X(x_5)}:>}
 \  \non \\&&\times{<:S_{\al}(x_4):S_{\be}(x_5):\psi^i(x_1):>},\nonumber\\
I_2&=&{<:e^{\alpha' ik_1.X(x_1)}:e^{\alpha' ik_2.X(x_2)}
:\partial X^b(x_3)e^{\alpha' ik_3.X(x_3)}:e^{i\frac{\alpha'}{2}p.X(x_4)}:e^{i\frac{\alpha'}{2}p.D.X(x_5)}:>}
 \  \non \\&&\times{<:S_{\al}(x_4):S_{\be}(x_5)::\psi^i(x_1):\alpha' ik_2.\psi\psi^{a}(x_2)>},\nonumber\\
 I_3&=&{<: e^{\alpha' ik_1.X(x_1)}:\partial X^a(x_2)e^{\alpha' ik_2.X(x_2)}
:e^{\alpha' ik_3.X(x_3)}:e^{i\frac{\alpha'}{2}p.X(x_4)}:e^{i\frac{\alpha'}{2}p.D.X(x_5)}:>}
 \  \non \\&&\times{<:S_{\al}(x_4):S_{\be}(x_5)::\psi^i(x_1):\alpha' ik_3.\psi\psi^{b}(x_3)>},\nonumber\\
 I_4&=&{<: e^{\alpha' ik_1.X(x_1)}:e^{\alpha' ik_2.X(x_2)}
:e^{\alpha' ik_3.X(x_3)}:e^{i\frac{\alpha'}{2}p.X(x_4)}:e^{i\frac{\alpha'}{2}p.D.X(x_5)}:>}
 \  \non \\&&\times{<:S_{\al}(x_4):S_{\be}(x_5):\psi^i(x_1)
:\alpha' ik_{2}\cd\psi\psi^a(x_2):\alpha' ik_{3}\cd\psi\psi^b(x_3):>}.
\label{i1234}
\eeqa

\vskip 0.1in

These correlators can be computed in straightforward fashion using Wick's theorem.
To obtain correlation function between two spin operators and several fermion fields and currents see Appendix A of \cite{Hatefi:2012wj}.

 The correlation function between two spin fields, two currents
and one worldsheet fermion -which we call $I_6^{bdaci}$ - is more complicated:
\beqa
I_6^{bdaci}&=&<:S_{\al}(x_4):S_{\be}(x_5)::\psi^i(x_1):\psi^c\psi^a(x_2):\psi^d\psi^b(x_3)>\nonumber\\
&=&\bigg\{(\Gamma^{bdaci}C^{-1})_{{\alpha\beta}}+\alpha' r_1\frac{Re[x_{24}x_{35}]}{x_{23}x_{45}}+\alpha'^2 r_2\bigg(\frac{Re[x_{24}x_{35}]}{x_{23}x_{45}}\bigg)^{2}
\bigg\}\nonumber\\&&
2^{-5/2}x_{45}^{5/4}(x_{24}x_{25}x_{34}x_{35})^{-1}(x_{14}x_{15})^{-1/2},
\label{hh}
\eeqa
where
\beqa
r_1&=&\bigg(\eta^{cd}(\Gamma^{bai}C^{-1})_{\alpha\beta}
-\eta^{cb}(\Gamma^{dai}C^{-1})_{\alpha\beta}-\eta^{ad}(\Gamma^{bci}C^{-1})_{\alpha\beta}+\eta^{ab}(\Gamma^{dci}C^{-1})_{\alpha\beta}\bigg),\nonumber\\
r_2&=&\bigg((-\eta^{cd}\eta^{ab}+\eta^{ad}\eta^{cb})(\gamma^{i}C^{-1})_{\alpha\beta}\bigg)
\eeqa
Substituting these results, one finds, after some algebra,
\beqa
{\cal A}^{C\phi AA}&\!\!\!\!\sim\!\!\!\!\!&\int dx_{1}dx_{2} dx_{3}dx_{4}dx_{5}(P_{-}\fsH_{(n)}M_p)^{\al\be}I\xi_{1i}\xi_{2a}\xi_{3b}x_{45}^{-1/4}(x_{14}x_{15})^{-1/2}\nonumber\\&&\times
\bigg(I_7^i(-\eta^{ab}x_{23}^{-2}+a^a_1a^b_2)+a^a_1a^{bi}_3+a^b_2a^{ai}_4-\alpha'^2 k_{2c}k_{3d}I_6^{bdaci}\bigg)\Tr(\lam_1\lam_2\lam_3)
\labell{amp3q},\eeqa
where
\beqa
I&=&|x_{12}|^{\alpha'^2 k_1.k_2}|x_{13}|^{\alpha'^2 k_1.k_3}|x_{14}x_{15}|^{\frac{\alpha'^2}{2} k_1.p}|x_{23}|^{\alpha'^2 k_2.k_3}|
x_{24}x_{25}|^{\frac{\alpha'^2}{2} k_2.p}
|x_{34}x_{35}|^{\frac{\alpha'^2}{2} k_3.p}|x_{45}|^{\frac{\alpha'^2}{4}p.D.p},\nonumber\\
a^a_1&=&ik_1^{a}\bigg(\frac{x_{14}}{x_{12}x_{24}}+\frac{x_{15}}{x_{25}x_{12}}\bigg)
+ik_3^{a}\bigg(\frac{x_{43}}{x_{24}x_{23}}+\frac{x_{53}}{x_{25}x_{23}}\bigg),\nonumber\\
a^b_2&=&ik_1^{b}\bigg(\frac{x_{14}}{x_{13}x_{34}}+\frac{x_{15}}{x_{35}x_{13}}\bigg)
+ik_2^{b}\bigg(\frac{x_{24}}{x_{34}x_{23}}+\frac{x_{25}}{x_{35}x_{23}}\bigg),\nonumber\\
a^{bi}_3&=&\alpha' ik_{3d}2^{-3/2}x_{45}^{1/4}(x_{34}x_{35})^{-1}(x_{14}x_{15})^{-1/2} \bigg\{(\Gamma^{bdi}C^{-1})_{\alpha\beta}\bigg\} ,\nonumber\\
a^{ai}_4&=&\alpha' ik_{2c}2^{-3/2}x_{45}^{1/4}(x_{24}x_{25})^{-1}(x_{14}x_{15})^{-1/2} \bigg\{(\Gamma^{aci}C^{-1})_{\alpha\beta}\bigg\}
,\nonumber\\
I_7^i&=&<:S_{\al}(x_4):S_{\be}(x_5):\psi^i(x_1):>=2^{-1/2}x_{45}^{-3/4}(x_{14}x_{15})^{-1/2}
(\gamma^{i}C^{-1})_{\alpha\beta}.\nonumber
\eeqa

\vskip 0.2in
Explicitly evaluating the integrals over the closed string location(to deal with the integrals and to provide more details see Appendix B of \cite{Hatefi:2012wj} ),
one can write the amplitude (\ref{amp3q}) as
\beqa {\cal A}^{\phi A A C}&=&{\cal A}_{1}+{\cal A}_{2}+{\cal A}_{3}+{\cal A}_{4}+{\cal A}_{5}\labell{711u}\eeqa
where
\beqa
{\cal A}_{1}&\!\!\!\sim\!\!\!&-2^{-1/2}\xi_{1i}\xi_{2a}\xi_{3b}
\bigg[k_{3d}k_{2c}\Tr(P_{-}\fsH_{(n)}M_p\Gamma^{bdaci})
\bigg]
L_1,
\nonumber\\
{\cal A}_{2}&\sim&2^{-1/2}\Tr(P_{-}\fsH_{(n)}M_p \Gamma^{bdi})\xi_{1i}\xi_{3b}k_{3d}
\bigg\{2k_1.\xi_2 L_2-2k_3.\xi_2 L_5
\bigg\}\nonumber\\
{\cal A}_{3}&\sim&-2^{-1/2}\Tr(P_{-}\fsH_{(n)}M_p \Gamma^{aci})\xi_{1i}\xi_{2a}k_{2c}
\bigg\{-2k_2.\xi_3 L_5+2k_1.\xi_3 L_3
\bigg\}\\
{\cal A}_{4}&\sim&-2^{-1/2}L_5
\bigg\{ \xi_{3b}\xi_{1i}\xi_{2a}\Tr(P_{-}\fsH_{(n)}M_p \Gamma^{bai})u +2k_2.\xi_3 k_{3d}\xi_{1i}\xi_{2a}\Tr(P_{-}\fsH_{(n)}M_p \Gamma^{dai})
\nonumber\\&&+2k_3.\xi_2 k_{2c}\xi_{1i}\xi_{3b}\Tr(P_{-}\fsH_{(n)}M_p \Gamma^{bci})-\Tr(P_{-}\fsH_{(n)}M_p \Gamma^{dci})k_{3d}k_{2c}\xi_{1i}(2\xi_2.\xi_3)
\bigg\}\nonumber\\
{\cal A}_{5}&\sim&2^{-1/2}\Tr(P_{-}\fsH_{(n)}M_p \gamma^{i})\xi_{1i}
\bigg\{
\xi_{3}.\xi_{2}(2ts)+2k_1.\xi_3(2k_3.\xi_2)t-4u k_1.\xi_2(k_1.\xi_3)+4sk_2.\xi_3k_1.\xi_2\bigg\}L_6\nonumber
\labell{483}\eeqa
where the functions
 $L_1,L_2,L_3,L_5,L_6$ are
\beqa
L_1&=&(2)^{-2(t+s+u)+1}\pi{\frac{\Gamma(-u+\frac{1}{2})
\Gamma(-s+\frac{1}{2})\Gamma(-t+\frac{1}{2})\Gamma(-t-s-u+1)}
{\Gamma(-u-t+1)\Gamma(-t-s+1)\Gamma(-s-u+1)}},\nonumber\\
L_2&=&(2)^{-2(t+s+u)}\pi{\frac{\Gamma(-u+1)
\Gamma(-s+1)\Gamma(-t)\Gamma(-t-s-u+\frac{1}{2})}
{\Gamma(-u-t+1)\Gamma(-t-s+1)\Gamma(-s-u+1)}}
\nonumber\\
L_3&=&(2)^{-2(t+s+u)}\pi{\frac{\Gamma(-u+1)
\Gamma(-s)\Gamma(-t+1)\Gamma(-t-s-u+\frac{1}{2})}{\Gamma(-u-t+1)
\Gamma(-t-s+1)\Gamma(-s-u+1)}}
,\nonumber\\
L_5&=&(2)^{-2(t+s+u)}\pi{\frac{\Gamma(-u)
\Gamma(-s+1)\Gamma(-t+1)\Gamma(-t-s-u+\frac{1}{2})}
{\Gamma(-u-t+1)\Gamma(-t-s+1)\Gamma(-s-u+1)}}
,\nonumber\\
L_6&=&(2)^{-2(t+s+u)-1}\pi{\frac{\Gamma(-u+\frac{1}{2})
\Gamma(-s+\frac{1}{2})\Gamma(-t+\frac{1}{2})\Gamma(-t-s-u)}
{\Gamma(-u-t+1)\Gamma(-t-s+1)\Gamma(-s-u+1)}},
\label{Ls}
\eeqa
It is possible to further simplify the result above:
\beqa {\cal A}^{\phi AA C}&=&{\hat{\cal A}}_{1}+\hat{{\cal A}}_{2}
+\hat{{\cal A}}_{3}+\hat{{\cal A}}_{4},\labell{141u}\eeqa
where
\beqa
\hat{{\cal A}}_{1}&\!\!\!\sim\!\!\!& 2^{1/2}\xi_{1i}\xi_{2a}\xi_{3b}
\bigg[k_{3d}k_{2c}\Tr(P_{-}\fsH_{(n)}M_p\Gamma^{bdaci})
\bigg]
(t+s+u)\hat{L}_1,
\nonumber\\
\hat{{\cal A}}_{2}&\sim&2^{-1/2}\Tr(P_{-}\fsH_{(n)}M_p \Gamma^{bdi})\xi_{1i}
\bigg\{2k_1.\xi_2 k_{3d}\xi_{3b}\hat{L}_2+2k_3.\xi_2 \xi_{3b} k_{1d}\hat{L}_5
-[2\leftrightarrow 3]
\bigg\} \\
\hat{{\cal A}}_{3}&\sim&-2^{-1/2}\hat{L}_5
\bigg\{ \xi_{3b}\xi_{1i}\xi_{2a}\Tr(P_{-}\fsH_{(n)}M_p \Gamma^{bai})u -\Tr(P_{-}\fsH_{(n)}M_p \Gamma^{dci})k_{3d}k_{2c}\xi_{1i}(2\xi_2.\xi_3)
\bigg\}\nonumber\\
\hat{{\cal A}}_{4}&\sim&2^{-3/2}\Tr(P_{-}\fsH_{(n)}M_p \gamma^{i})\xi_{1i}
\bigg\{
\xi_{3}.\xi_{2}(2ts)+2k_1.\xi_3(2k_3.\xi_2)t-4u k_1.\xi_2(k_1.\xi_3)+4sk_2.\xi_3k_1.\xi_2\bigg\}\hat{L}_1\nonumber
\labell{4798}
\eeqa
The functions $\hat{L}_1,\hat{L}_2$ are ($\hat{L_5}=L_5$)
\beqa
\hat{L}_1&=&(2)^{-2(t+s+u)}\pi{\frac{\Gamma(-u+\frac{1}{2})
\Gamma(-s+\frac{1}{2})\Gamma(-t+\frac{1}{2})\Gamma(-t-s-u)}
{\Gamma(-u-t+1)\Gamma(-t-s+1)\Gamma(-s-u+1)}},\nonumber\\
\hat{L}_2&=&(2)^{-2(t+s+u)}\pi{\frac{\Gamma(-u+1)
\Gamma(-s+1)\Gamma(-t)\Gamma(-t-s-u+\frac{1}{2})}
{\Gamma(-u-t+1)\Gamma(-t-s+1)\Gamma(-s-u+1)}}
\nonumber\\
\label{Ls}
\eeqa

The amplitude satisfies ward identities as it should:  by replacing
$\xi_{2a}\rightarrow k_{2a}$ and $\xi_{3b}\rightarrow k_{3b}$, it vanishes.
As in previous works, the amplitude is non-vanishing for certain values of $p$ and $n$: $n=p-2,  n=p+2$ and $p=n$.
The amplitude has an infinite number of massless scalar and gauge field poles and in addition it has infinite contact interactions.
We must expand it at low energy limit where the momentum expansions have been introduced in detail in \cite{Hatefi:2010ik}.

\vskip 0.2in

Expansion of the functions $\hat{L}_1,\hat{L}_2$
around the $t,s,u \rightarrow 0$ is
\beqa
\hat{L}_1 &=&-{\pi^{5/2}}\left( \sum_{n=0}^{\infty}c_n(s+t+u)^n\right.
\left.+\frac{\sum_{n,m=0}^{\infty}c_{n,m}[s^n t^m +s^m t^n]}{(t+s+u)}\right.\nonumber\\
&&\left.+\sum_{p,n,m=0}^{\infty}f_{p,n,m}(s+t+u)^p[(s+t)^{n}(st)^{m}]\right)\nonumber\\
\hat{L}_2 &=&-\pi^{3/2}\sum_{n=-1}^{\infty}b_n \frac{1}{t}(u+s)^{n+1}+\sum_{p,n,m=0}^{\infty}e_{p,n,m}t^{p}(su)^{n}(s+u)^m
\labell{highcaap}.
\eeqa

where some of the coefficients $b_n,\,e_{p,n,m},\,c_n,\,c_{n,m}$ and $f_{p,n,m}$ are
\beqa
&&b_{-1}=1,\,b_0=0,\,b_1=\frac{1}{6}\pi^2,\,b_2=2\z(3),c_0=0,c_1=-\frac{\pi^2}{6},\nonumber\\
&&e_{2,0,0}=e_{0,1,0}=2\z(3),e_{1,0,0}=\frac{1}{6}\pi^2,e_{1,0,2}=\frac{19}{60}\pi^4,e_{1,0,1}=e_{0,0,2}=6\z(3),\nonumber\\
&&e_{0,0,1}=\frac{1}{3}\pi^2,e_{3,0,0}=\frac{19}{360}\pi^4,e_{0,0,3}=e_{2,0,1}=\frac{19}{90}\pi^4,e_{1,1,0}=e_{0,1,1}=\frac{1}{30}\pi^4,\labell{577}\\
&&c_2=-2\z(3),
\,c_{1,1}=\frac{\pi^2}{6},\,c_{0,0}=\frac{1}{2},c_{3,1}=c_{1,3}=\frac{2}{15}\pi^4,c_{2,2}=\frac{1}{5}\pi^4,\nonumber\\
&&c_{1,0}=c_{0,1}=0,
c_{3,0}=c_{0,3}=0\,
,\,c_{2,0}=c_{0,2}=\frac{\pi^2}{6},c_{1,2}=c_{2,1}=-4\z(3),\nonumber\\
&&f_{0,1,0}=\frac{\pi^2}{3},\,f_{0,2,0}=-f_{1,1,0}=-6\z(3),f_{0,0,1}=-2\z(3),c_{4,0}=c_{0,4}=\frac{1}{15}\pi^4.\, \nonumber
\eeqa

Note that the coefficients $b_n$ are exactly the coefficients that appear in the momentum expansion of the S-matrix element of
$CAAA$
\cite{Hatefi:2010ik}.
 $c_n,\,c_{n,m},f_{p,n,m}$ are different from those coefficients that appeared in \cite {Hatefi:2008ab}.

\section{Two scalar and two  gauge field couplings }

The closed form result of  $<V_{RR}V_{\phi} V_A V_A>$ above can be
momentum-expanded to be compared with the corresponding effective field theory computations.
In the earlier work \cite{Hatefi:2010ik}, it was shown that the string
theory result of $<V_{RR}V_A V_A V_A>$ is reproduced by the following
SYM vertices
\beqa
-T_p (2\pi\alpha')^4 S\Tr\left(-\frac{1}{8}F_{bd}F^{df}F_{fh}F^{hb}+\frac{1}{32}(F_{ab}F^{ba})^2\right).
\label{55}
\eeqa
Extension of these interaction vertices to higher derivative
couplings reproduced both the massless pole terms and
the contact terms.

Below, we will show that the same is true for the couplings two scalar fields and two gauge fields.
 First, we show that the correct couplings between two gauge fields and two scalar fields are
\beqa
&&- \frac{T_p(2\pi\alpha')^4}{2}{\rm STr}
\left(D_a\phi^iD^b\phi_iF^{ac}F_{bc}-\frac{1}{4}
(D_a\phi^i D^a\phi_iF^{bc}F_{bc})\right).\labell{a011}
\eeqa
In order to find the infinite higher derivative corrections to two scalar fields and two
gauge fields, one must
find the amplitude of two scalars and two gauge fields.
Computing this amplitude and working out details
, we could find the higher derivative
corrections of two scalars and two gauge fields to all orders of $\alpha'$ as the following:
\beqa
(2\pi\alpha')^4\frac{1}{ 2 \pi^2}T_p\left(\alpha'\right)^{n+m}\sum_{m,n=0}^{\infty}(\cL_{1}^{nm}+\cL_{2}^{nm}+\cL_{3}^{nm}),\labell{highder}\eeqa
\beqa
&&\cL_{1}^{nm}=-
\Tr\left(\frac{}{}a_{n,m}\cD_{nm}[D_a \phi^i D^b \phi_i F^{ac}F_{bc}]+ b_{n,m}\cD'_{nm}[D_a \phi^i F^{ac} D^b \phi_i F_{bc}]+h.c.\frac{}{}\right),\nonumber\\
&&\cL_{2}^{nm}=-\Tr\left(\frac{}{}a_{n,m}\cD_{nm}[D_a \phi^i D^b \phi_i F_{bc}F^{ac}]+\frac{}{}b_{n,m}\cD'_{nm}[D_a \phi^i F_{bc} D^b \phi_i F^{ac}]+h.c.\frac{}{}\right),\nonumber\\
&&\cL_{3}^{nm}=\frac{1}{2}\Tr\left(\frac{}{}a_{n,m}\cD_{nm}[D_a \phi^i D^a \phi_i F^{bc}F_{bc}]+\frac{}{}b_{n,m}\cD'_{nm}[D_a \phi^i F_{bc} D^a \phi_i F^{bc}]+h.c\frac{}{}\right),\nonumber\eeqa
where the higher derivative operators
$D_{nm} $ and $ D'_{nm}$ are defined \cite{Hatefi:2010ik} as
\beqa
\cD_{nm}(EFGH)&\equiv&D_{b_1}\cdots D_{b_m}D_{a_1}\cdots D_{a_n}E  F D^{a_1}\cdots D^{a_n}GD^{b_1}\cdots D^{b_m}H,\nonumber\\
\cD'_{nm}(EFGH)&\equiv&D_{b_1}\cdots D_{b_m}D_{a_1}\cdots D_{a_n}E   D^{a_1}\cdots D^{a_n}F G D^{b_1}\cdots D^{b_m}H.\nonumber
\eeqa
As usual, the above couplings are valid up to total derivative
terms and terms such as $\prt_a\prt^aFFD\phi D\phi$ that vanish on-shell.
 They have no effect on the massless poles of S-matrix elements,
because by canceling $k^2$ with the massless propagator, they produce
 contact terms.
 We now turn to verification of (\ref{highder}) and the terms
coming from the DBI part.

\subsection{Infinite number of massless scalar poles for $p+2=n$ case }

In this subsection, we check that the two gauge two scalar
interaction vertices \reef{highder} produce an infinite number of massless scalar poles of the string theory S-matrix element that are in the $(s+t+u)$-channel.
Specifically, the goal is to show that the massless poles of
the string computation are reproduced by the following
WZ coupling (that was found in \cite{Garousi:2000ea}),
\beqa
\lambda\mu_p\int d^{p+1}\sigma {1\over (p+1)!}
(\veps^v)^{a_0\cdots a_{p}}\,\Tr\left(\phi^i\right)\,
H^{(p+2)}_{ia_0\cdots a_{p}}(\sigma) \eeqa
and by the higher derivative two gauge field and two scalar couplings that have been found in \reef{highder}. To that end, let us consider the amplitude of the decay of one R-R field to one scalar and two gauge fields in the world volume theory of the BPS branes. In the (Feynman gauge) Feynman diagrammatic rules, it is given by
\beqa
{\cal A}&=&V_{\alpha}^{i}(C_{p+1},\phi)G_{\alpha\beta}^{ij}(\phi)V_{\beta}^{j}(\phi,\phi_1,
A_2,A_3),\labell{amp549}
\eeqa
where
\beqa
G_{\alpha\beta}^{ij}(\phi) &=&\frac{-i\delta_{\alpha\beta}\delta^{ij}}{T_p(2\pi\alpha')^2
k^2}=\frac{-i\delta_{\alpha\beta}\delta^{ij}}{T_p(2\pi\alpha')^2
(t+s+u)},\nonumber\\
V_{\alpha}^{i}(C_{p+1},\phi)&=&i(2\pi\alpha')\mu_p\frac{1}{(p+1)!}(\veps^v)^{a_0\cdots a_{p}}
 H^{i(p+2)}_{a_0\cdots a_{p}}\Tr(\lambda_{\alpha}).
\labell{Fey}
\eeqa
We have replaced $k^2$
 by $(t+s+u)$ in the first equation of (\ref{Fey}). In the second equation of  (\ref{Fey}), $\Tr(\lambda_{\alpha})$ is non-zero for the abelian matrix $\lambda_{\alpha}$.
Noting the fact that the off-shell scalar field must be abelian, one finds the higher derivative vertex  $ V_{\beta}^{j}(\phi,\phi_1,
A_2,A_3)$  from the higher derivative couplings
in \reef{highder}:
\beqa
V_{\beta}^{j}(\phi,\phi_1,
A_2,A_3)&=&\xi_{1}^j\frac{I_8}{2\pi^2}(\alpha')^{n+m}(a_{n,m}+b_{n,m})
\bigg(\frac{}{}(k_3\inn k_1)^m(k_1\inn k_2)^n+(k_3\inn k)^m(k_2\inn k)^n
\nonumber\\&&+(k_1\inn k_3)^n(k_1\inn k_2)^m+(k\inn k_3)^n (k\inn k_2)^m
\bigg),\labell{verppaa}\eeqa
where  $k$ is the momentum of the off-shell scalar field, and
\beqa
I_8&=&(2\pi\alpha')^4T_{p}\Tr(\lam_1\lam_2\lam_3\lambda_{\beta})\bigg[\frac{st}{2}\xi_{2}.\xi_3+t k_1.\xi_3 k_3.\xi_2+s k_1.\xi_2 k_2.\xi_3-u k_1.\xi_2 k_1.\xi_3\bigg]\eeqa
For the reason explained in \cite{Hatefi:2010ik}, $b_{n,m}$ is symmetric.
 Note that we must consider two permutations,
\beqa
\Tr(\lam_1\lambda_{\beta}\lam_2\lam_3),
\Tr(\lambda_{\beta}\lam_1\lam_2\lam_3)
\eeqa
to obtain the desired 123 ordering
of the amplitude. Let us list some of the coefficients $a_{n,m}$
and $b_{n,m}$ for convenience:
\beqa
&&a_{0,0}=-\frac{\pi^2}{6},\,b_{0,0}=-\frac{\pi^2}{12},a_{1,0}=2\z(3),\,a_{0,1}=0,\,b_{0,1}=-\z(3),a_{1,1}=a_{0,2}=-7\pi^4/90,\nonumber\\
&&a_{2,2}=(-83\pi^6-7560\z(3)^2)/945,b_{2,2}=-(23\pi^6-15120\z(3)^2)/1890,a_{1,3}=-62\pi^6/945,\nonumber\\
&&\,a_{2,0}=-4\pi^4/90,\,b_{1,1}=-\pi^4/180,\,b_{0,2}=-\pi^4/45,a_{0,4}=-31\pi^6/945,a_{4,0}=-16\pi^6/945,\nonumber\\
&&a_{1,2}=a_{2,1}=8\z(5)+4\pi^2\z(3)/3,\,a_{0,3}=0,\,a_{3,0}=8\z(5),b_{1,3}=-(12\pi^6-7560\z(3)^2)/1890,\nonumber\\
&&a_{3,1}=(-52\pi^6-7560\z(3)^2)/945, b_{0,3}=-4\z(5),\,b_{1,2}=-8\z(5)+2\pi^2\z(3)/3,\nonumber\\
&&b_{0,4}=-16\pi^6/1890.\eeqa
(They were computed in \cite{Hatefi:2010ik}
for the four gauge field amplitude. In retrospect, it must be due to field theory T-duality
that they remain valid for the present two scalars  and two gauge fields amplitude.)
Now one can write  $k_3\inn k=k_2.k_1-(k^2)/2$ and $k_2\inn k=k_1.k_3-(k^2)/2$.
The terms $k^2$ in the vertex \reef{verppaa} will be cancelled with the $k^2$ in the denominator of the scalar field propagator producing
contact terms. They will not be explicitly considered in this work. Setting them aside, one finds the following result of an infinite number of massless poles,
\beqa
&&16\pi\mu_p\frac{\eps^{a_{0}\cdots a_{p}}\xi_1^i
H^{i(p+2)}_{a_0\cdots a_{p}}}{(p+1)!(s+t+u)}\Tr(\lam_1\lam_2\lam_3)
\sum_{n,m=0}^{\infty}\bigg((a_{n,m}+b_{n,m})[s^{m}t^{n}+s^{n}t^{m}]\nonumber\\&&
\bigg[2st\xi_{2}.\xi_3+4t k_1.\xi_3 k_3.\xi_2+4s k_1.\xi_2 k_2.\xi_3-4u k_1.\xi_2 k_1.\xi_3\bigg]
\label{amphigh8}\eeqa
\\
As a check of our calculations let us compare the above amplitude with the infinite  number of massless poles in the string theory result. We will take several
 values of $n,m$ for illustrations. Common factors of the
string and field theory amplitudes will be
omitted. For $n=m=0$, the amplitude \reef{amphigh8} has the following numerical factor
\beqa
-2(a_{0,0}+b_{0,0})&=&-2(\frac{-\pi^2}{6}+\frac{-\pi^2}{12})=\frac{\pi^2}{2}\nonumber\eeqa
There is a corresponding term in the string amplitude, and it has the numerical factor  $(2\frac{\pi^2}{2}c_{0,0})$ which is equal to the number above.  At the order of $\alpha'$, the amplitude \reef{amphigh8} has the following numerical factor
\beqa
-(a_{1,0}+a_{0,1}+b_{1,0}+b_{0,1})(s+t)&=&0\nonumber\eeqa
The corresponding term in string amplitude is proportional to   $\frac{\pi^2}{2}(c_{1,0}+c_{0,1})(s+t)$ which indeed vanishes.  At the  order of $(\alpha')^2$, the amplitude \reef{amphigh8} has the following factor
\beqa
&&-2(a_{1,1}+b_{1,1})st-(a_{0,2}+a_{2,0}+b_{0,2}+b_{2,0})[s^2+t^2]\nonumber\\
&&=\frac{\pi^4}{6}(st)+\frac{\pi^4}{6}(s^2+t^2)
\nonumber\eeqa
The string result has the numerical factor  $\frac{\pi^2}{2}[c_{1,1}(2st)+(c_{2,0}+c_{0,2})(s^2+t^2)]$, which is equal to the above factor using the coefficients in \reef{577}.
At the order of $\alpha'^3$, this amplitude has the following factor
\beqa
&&- (a_{3,0}+a_{0,3}+b_{0,3}+b_{3,0})[s^3+t^3]- (a_{1,2}+a_{2,1}+b_{1,2}+b_{2,1})[st(s+t)]\nonumber\\
&&=-4\pi^2\z(3)st(s+t)
\nonumber\eeqa
which is equal to the corresponding term  , \ie $\frac{\pi^2}{2}[(c_{0,3}+c_{3,0})[s^3+t^3]+(c_{2,1}+c_{1,2})st(s+t)]$.  
At the  order of $(\alpha')^4$, the amplitude \reef{amphigh8} has the following factor
\beqa
&&- (a_{4,0}+a_{0,4}+b_{0,4}+b_{4,0})(s^4+t^4)- (a_{3,1}+a_{1,3}+b_{3,1}+b_{1,3})[st(s^2+t^2)]\nonumber\\
&&-2(a_{2,2}+b_{2,2}) s^2t^2=\frac{\pi^6}{15}(s^4+t^4+2(s^3t+t^3s)+3s^2t^2)
\nonumber\eeqa
The string result has the numerical factor  $\frac{\pi^2}{2}[(c_{4,0}+c_{0,4})(s^4+t^4)+(c_{1,3}+c_{3,1})(s^3t+t^3s)+2c_{2,2} s^2t^2]$ which is equal to the above factor using the coefficients in \reef{577}. Note that  the string amplitude has been rescaled by  $2^{1/2}\pi^{1/2}\mu_p$.

Similar comparisons can be carried out\footnote{The same checks have been done for finding an infinite number of massless poles of $<V_CV_AV_AV_A>$ in \cite{Hatefi:2010ik}.
} for all higher orders
of $\alpha'$: the field theory amplitude \reef{amphigh8} exactly reproduces the infinite number of massless scalar poles of the string theory amplitude of $<V_CV_\phi V_AV_A>$.
 This shows that the higher derivative couplings of two scalar two gauge fields
are exact up to terms that vanish on-shell. They are also consistent with the momentum expansion of the amplitude of $AA\phi\phi$.

\subsection{Infinite number of massless scalar  poles for $p=n$ case }

Substituting the expansion of $\hat{L}_2$ mentioned in \reef{highcaap} into the amplitude, it is possible to obtain all massless scalar poles in the string theory side:
\beqa
\hat{{\cal A}}_{2}&\sim&2^{-1/2}\Tr(P_{-}\fsH_{(n)}M_p \Gamma^{bdi})\xi_{1i}
\bigg\{2k_1.\xi_2 \xi_{3b} k_{3d}L_2-[2\leftrightarrow 3]
\bigg\}
\eeqa
 The trace can be calculated straightforwardly:
\beqa
\Tr\bigg(\fsH_{(n)}M_p\Gamma^{bdi}
\bigg)&=&\pm\frac{32}{p!}\eps^{a_{0}\cdots a_{p-2}bd}H^{i}_{a_{0}\cdots a_{p-2}},
\nonumber\eeqa
Substituting it into the amplitude and keeping
all the scalar poles, one gets
\beqa
A^{C\phi AA}&=&\mp\frac{32}{2 p!}\mu_p \pi^2\xi_{1i}\eps^{a_{0}\cdots a_{p-2}bd}H^{i}_{a_{0}\cdots a_{p-2}}\nonumber\\&&\bigg\{\bigg[\sum_{n=-1}^{\infty}\frac{1}{t}{b_n(u+s)^{n+1}}(2k_1.\xi_2)\xi_{3b}k_{3d}\bigg]
-\bigg[2\leftrightarrow 3\bigg]
\bigg\}\Tr(\lam_1\lam_2\lam_3).\label{UI}
\eeqa
where the amplitude is rescaled by $2^{1/2}\pi^{1/2}\mu_{p}$. The amplitude is antisymmetric under the interchange of the gauge fields; therefore, the whole amplitude vanishes for the abelian gauge group. The amplitude satisfies the Ward identity under $\xi_{3b} \rightarrow k_{3b}$. Note that we kept only
the massless poles; the other terms are
contact terms infinite in number. Let us analyze all
orders of the massless scalar poles.

Because the amplitudes in s and t-channels are similar, we
analyze the amplitude in the t-channel in detail; by interchanging the labels of momentum and polarization vectors, $(2\leftrightarrow 3)$, the other massless poles in the s channel can easily be obtained. The relevant vertices
for this case in the field theory side is
\beqa
S^{(1)}&=&i\lambda\mu_p\int \STr\left(F P\left[
C^{(p-1)}(\s,\phi)\right]\right)
\labell{interac}\\
&=&{i}\lambda^2\mu_p\int d^{p+1}\s {1\over p!}(\veps^v)^{a_0\cdots a_{p}}
\left[{p}\Tr\left(F_{a_{0}a_{1}}\phi^k\right)
\prt_kC^{(p-1)}_{a_2\cdots a_{p}}(\s)\right]\,\,\, .
\nonumber
\eeqa
Where the scalar field comes from Taylor expansion (see section 5 of \cite{Hatefi:2012wj}). With this vertex, the massless poles in the t-channel should be reproduced as
\beqa
{\cal A}&=&V^i_{\alpha}(C_{p-1},A_3,\phi)G^{ij}_{\alpha\beta}(\phi)V^j_{\beta}(\phi,A_2,\phi_1),\labell{amp642}
\eeqa
The vertices and propagator above are
\beqa
V^i_{\alpha}(C_{p-1},A_3,\phi)&=&\frac{iN\lambda^2\mu_p}{(p)!}(\veps^v)^{a_0\cdots a_p}(H^{(p)})^i{}_{a_2\cdots a_p}\xi_{3a_{1}}k_{3a_{0}}\Tr(\lam_3\lambda_\alpha)\sum_{n=-1}^{\infty}b_n(\alpha'k_3.k)^{n+1},\nonumber\\
\eeqa
with
\beqa
({V}^{\phi\phi_1 A_2})^j_{\beta}&=&-2i\lambda^2T_p\Tr
(\xi_2\inn k_1[\xi_1^j,T_{\beta}])
\labell{fvertex}\\
({G}^{\phi})^{ij}_{\alpha\beta}&=&-\frac{i}{N\lambda^2T_p}\frac{\delta^{ij}
\delta_{\alpha\beta}}{k^2}\,\,\, ,
\nonumber
\eeqa
where $k$ is the momentum of the off-shell scalar field and it
satisfies
 $k^2=(k_1+k_2)^2=-t$. We also have written $\z^i={\z}^i_{\alpha}T_{\alpha}$
where $T_{\alpha}$ are the $U(N)$ generators with normalization
$\Tr(T_{\alpha}T_{\beta})=N\delta_{\alpha\beta}$.
The propagator is derived from the standard gauge kinetic term resulting in the expansion of the Born-Infeld action. The vertex $V^j_{\beta}(\phi,A_2,\phi_1)$ has found from the standard non-abelian kinetic term of
the scalar field; the vertex $V^i_{\alpha}(C_{p-1},A_3,\phi)$ is found from the higher derivative extension of the WZ coupling $\Tr(\partial_{k}C_{p-1}\wedge F\phi^k)$. The important point is that the vertex $V^j_{\beta}(\phi,A_2,\phi_1)$ has no higher derivative correction as it comes from the kinetic term of the scalar field.

\vskip 0.1in

Substituting all vertices in the amplitude \reef{amp642}, one finds
\beqa
{\cal A}&=&\mu_p(2\pi\alpha')^{2}\frac{1}{(p)!t}\Tr(\lam_1\lam_2\lam_3)\eps^{a_{0}\cdots a_{p-2}ba}H^{i}_{a_{0}\cdots a_{p-2}}\sum_{n=-1}^{\infty}b_n\bigg(\frac{\alpha'}{2}\bigg)^{n+1}(s+u)^{n+1}\nonumber\\&&\times\bigg[-2k_1.\xi_2\xi_{1i}\xi_{3b} k_{3a}\bigg]\eeqa
It describes exactly the same infinite number of massless poles of the string theory amplitude \reef{UI} in t-channel; there is precise agreement between the field theory calculation and string result.
A similar comparison in $s$-channel also yields agreement.

\subsection{Infinite number of massless gauge field poles for $p=n$ case }

The expansion of the function $\hat{L}_5$ is given by
\beqa
\hat{L}_5&=&-\pi^{3/2}\bigg(\frac{1}{u}\sum_{n=-1}^{\infty}b_n(t+s)^{n+1}+\sum_{p,n,m=0}^{\infty}e_{p,n,m}u^{p}(st)^{n}(s+t)^m\bigg).\labell{high}\eeqa
One can read off the infinite massless gauge field poles for the amplitude of $<V_CV_\phi V_A V_A>$:
\beqa
\hat{{\cal A}}_{2}&\sim&2^{-1/2}L_5\bigg(\Tr(P_{-}\fsH_{(n)}M_p \Gamma^{bdi})\xi_{1i}\xi_{3b}k_{3d}
(-2k_3.\xi_2 )-\Tr(P_{-}\fsH_{(n)}M_p \Gamma^{aci})\xi_{1i}\xi_{2a}k_{2c}
(-2k_2.\xi_3 )\nonumber\\&&
-2k_2.\xi_3 k_{3d}\xi_{1i}\xi_{2a}\Tr(P_{-}\fsH_{(n)}M_p \Gamma^{dai})
-2k_3.\xi_2 k_{2c}\xi_{1i}\xi_{3b}\Tr(P_{-}\fsH_{(n)}M_p \Gamma^{bci})\nonumber\\&&+\Tr(P_{-}\fsH_{(n)}M_p \Gamma^{dci})k_{3d}k_{2c}\xi_{1i}(2\xi_2.\xi_3)\bigg)
\eeqa
Working out the trace, one finds the infinite
gauge field poles
\beqa
A^{C\phi AA}&=&\mp\frac{32}{2 p!}\mu_p \pi^2\xi_{1i}\eps^{a_{0}\cdots a_{p-2}dc}H^{i}_{a_{0}\cdots a_{p-2}}\Tr(\lam_1\lam_2\lam_3)\sum_{n=-1}^{\infty}\frac{1}{u}{b_n(t+s)^{n+1}}\bigg\{\bigg[2\xi_2.\xi_3 k_{3d}k_{2c}\nonumber\\&&-2k_3.\xi_2 \xi_{3d}(k_{2c}+k_{3c})+2k_2.\xi_3 \xi_{2d}(k_{2c}+k_{3c}) \bigg]
\bigg\}.\label{UIiop}\eeqa
The amplitude is antisymmetric under the interchange of the gauge
fields; it vanishes for the abelian gauge group.
 We have kept the entire massless poles; the
other terms are infinite contact terms.
Let us  focus on the massless gauge field poles. (A similar analysis can
be done for the contact terms.)
The relevant WZ coupling is
\beqa
S^{(2)}&=&i\lambda\mu_p\int \STr\left(F P\left[
C^{(p-1)}(\s,\phi)\right]\right)
\labell{interac}\\
&=&{i}\lambda^2\mu_p\int d^{p+1}\s {1\over p!}(\veps^v)^{a_0\cdots a_{p}}
\left[{p} \Tr\left(F_{a_{0}a_{1}}\phi^k\right)
\prt_kC^{(p-1)}_{a_2\cdots a_{p}}(\s)\right]\,\,\, .
\nonumber
\eeqa
An infinite number of massless poles in the u-channel should be reproduced in field theory according to the following Feynman rule;
the $u$-channel amplitude can be written as
\beqa
{\cal A}&=&V^a_{\alpha}(C_{p-1},\phi_1,A)G^{ab}_{\alpha\beta}(A)V^b_{\beta}(A,A_2,A_3),\labell{amp60942}\eeqa
The vertices and propagator can
be read from the effective action
 are
\beqa
V^a_{\alpha}(C_{p-1},\phi_1,A)&=&\frac{i\lambda^2\mu_p}{(p)!}(\veps^v)^{a_0\cdots a_{p-1}a}(H^{(p)})^k{}_{a_1\cdots a_{p-1}}\xi_{1k}k_{a_{0}}\Tr(\lambda_\alpha\lam_1)\sum_{n=-1}^{\infty}b_n(\alpha'k_1.k)^{n+1},\nonumber\\
\eeqa
with
\beqa
V^b_{\beta}(A,A_2,A_3)&=&-iT_p(2\pi\alpha')^{2}\Tr(\lam_2\lam_3\lambda_\beta)\bigg[\xi_2^b(k_2-k).\xi_3+
\xi_3^b(k-k_3).\xi_2+\xi_3.\xi_2(k_3-k_2)^b\bigg],\nonumber\\
G_{\alpha\beta}^{ab}(A)&=&\frac{i\delta_{\alpha\beta}\delta^{ab}}{(2\pi\alpha')^2 T_p
(k)^2},\nonumber\eeqa
where  $k$ is the momentum of the off-shell gauge field
and satisfies $k^2=(k_2+k_3)^2=-u$.
The vertex $V^a_{\alpha}(C_{p-1},\phi_1,A)$ has found from the higher derivative extension of the WZ coupling $\Tr(\partial_{k}C_{p-1}\wedge F\phi^k)$.
Substituting these vertices in the amplitude, one gets
\beqa
{\cal A}&=&\mu_p(2\pi\alpha')^{2}\frac{1}{(p)!u}\Tr(\lam_1\lam_2\lam_3)\eps^{a_{0}\cdots a_{p-2}bd}
(H^{(p)})^i{}_{a_0\cdots a_{p-2}}\sum_{n=-1}^{\infty}b_n\bigg(\frac{\alpha'}{2}\bigg)^{n+1}(s+t)^{n+1}\nonumber\\&&\times\bigg[2k_3.\xi_2\xi_{1i}\xi_{3b} k_{1d}-2k_2.\xi_3\xi_{1i}\xi_{2b} k_{1d}-2\xi_3.\xi_2\xi_{1i} k_{2b} k_{3d}\bigg]\eeqa
It again displays precise agreement with the infinite $u$-channel poles in
the string amplitude, \reef{UIiop}.



\section{Conclusion}

Following \cite{Hatefi:2010ik} in which a similar analysis was done
for $<V_{C}V_A V_AV_A>$, we have computed $<V_{C}V_{\phi} V_AV_A>$
in closed form. By performing the momentum expansion,
the corresponding low energy SYM vertices have
been determined in all orders in $\alpha'$. We believe that the result
of this paper will provide the basis for future research on, e.g.,
next-to-leading order dielectric effect and
other related topics in string theory \cite{Hatefi:2012sy}.

\vskip 0.1in

 For the simple scalar poles, since there is no correction to   $\Tr\left(\phi^i\right)\,
H^{i(p+2)}_{a_0\cdots a_{p}}$, the non-leading scalar poles should give  information about the higher derivative corrections to the couplings of two  scalars and two gauge fields where we found them up
 to all orders of $\alpha'$ in \reef{highder}.

\vskip 0.1in

Although we did not analyse the contact
terms of the string amplitude, they would give information about the higher derivative corrections
to $\Tr(\partial_{k}C_{p-3}\wedge F\wedge F\phi^k)$.
There are several other cases in which one could carry out
analogous analyses.

Another line of research
is associated with the subtlety brought up
 in the introduction with regards to any amplitude that contains
both open string vertex operators and closed string vertex
operators. In this work, we have followed a step that
seems to be the commonly implemented in literature, which we are
getting to now. Closed string coordinates have two sets of
oscillators; let us denote them by $\alpha_n$ and
$\tilde{\alpha}_n$
collectively. For an amplitude that involves
both open string states and closed string states, the
computations are typically done in a path integral setup where the
Green function is determined using conformal field
theory techniques.

This is for a good reason. Once one attempts the computations
in the oscillator, what to do with the second
set of a closed string oscillators, $\tilde{\alpha}_n$ is not clear in the framework of the first
quantized string.
The only viable option seems to be to use one set of
oscillators as was commented below (3.4) in
\cite{Billo:2006jm}. In other words, the two sets of
the closed string oscillators are identified with each other,
and in turn identified with the open string oscillators as well.\footnote{The
reason stated in \cite{Billo:2006jm} for the identification
was the presence of fractional branes. However, the same
practice seems to be adopted by other groups in later related works where fractional
branes are not present.}
As far as we can see, a certain "analytic continuation" is
involved in the prescription of identifying the
second set of the closed string oscillators as the first set.
We believe that there is a room for better understanding and
systematic study of this step.
In effect, the identification makes the closed string state
a composite state of the open string fields.
 At the effective field theory level,
this means that the supergravity background fields that
are present in the DBI action should become
composite, namely, they must be functions of the SYM fields.\footnote{
To determine the forms of the functions, that is, the proper
background, it would be necessary to rely on the full open string
setup and to go beyond the tree level: it is expected that quantum effects would play an important role as proposed in \cite{Park:2007mc}.}
Those background fields can then be "Taylor-expanded" as
discussed in the Myers' work \cite{Myers:1999ps}.

\newpage

\renewcommand{\theequation}{A.\arabic{equation}}
 \setcounter{equation}{0}
  \section*{Appendix A: Conventions
  }

 Our index conventions are such that lowercase
Greek indices take values in the whole ten-dimensional spacetime, e.g.,
\beqa
\mu,\nu = 0, 1,..., 9
\eeqa
 Early
Latin indices run along the world-volume,
\beqa
 a, b, c = 0, 1,..., p
\eeqa
while middle Latin indices
represent the transverse space
\beqa
i,j = p + 1,...,9.
\eeqa
 Doubling trick is implemented according to
\begin{displaymath}
\tilde{X}^{\mu}(\bar{z}) \rightarrow D^{\mu}_{\nu}X^{\nu}(\bar{z}) \ ,
\spa
\tilde{\psi}^{\mu}(\bar{z}) \rightarrow
D^{\mu}_{\nu}\psi^{\nu}(\bar{z}) \ ,
\spa
\tilde{\phi}(\bar{z}) \rightarrow \phi(\bar{z})\,, \mand
\tilde{S}_{\al}(\bar{z}) \rightarrow M_{\al}{}^{\be}{S}_{\be}(\bar{z})
 ,
\end{displaymath}
where
\begin{displaymath}
D = \left( \begin{array}{cc}
-1_{9-p} & 0 \\
0 & 1_{p+1}
\end{array}
\right) \ ,\,\, \mand
M_p = \left\{\begin{array}{cc}\frac{\pm i}{(p+1)!}\ga^{a_{1}}\ga^{a_{2}}\ldots \ga^{a_{p+1}}
\eps_{a_{1}\ldots a_{p+1}}\,\,\,\,{\rm for\, p \,even}\\ \frac{\pm 1}{(p+1)!}\ga^{a_{1}}\ga^{a_{2}}\ldots \ga^{a_{p+1}}\ga_{11}
\eps_{a_{1}\ldots a_{p+1}} \,\,\,\,{\rm for\, p \,odd}\end{array}\right.
\end{displaymath}
 The basic holomorphic correlators for the world-sheet fields $X^{\mu},\psi^{\mu}, \phi$ are
\begin{eqnarray}
\lan X^{\mu}(z)X^{\nu}(w)\ran & = & -\frac{\alpha'}{2}\eta^{\mu\nu}\log(z-w) , \non \\
\lan \psi^{\mu}(z)\psi^{\nu}(w) \ran & = & -\frac{\alpha'}{2}\eta^{\mu\nu}(z-w)^{-1} \ ,\non \\
\lan\phi(z)\phi(w)\ran & = & -\frac{\alpha'}{2}\log(z-w) \ .
\labell{prop}\end{eqnarray}
For convenience, we introduce
\beqa
 x_{4}\equiv\ z=x+iy\quad,\quad x_{5}\equiv\bz=x-iy
\eeqa
SL(2,R) symmetry is fixed
by choosing the positions of the open string vertex operators
 \beqa
 x_{1}=0 ,\qquad x_{2}=1,\qquad x_{3}\rightarrow \infty,
 \qquad dx_1dx_2dx_3\rightarrow x_3^{2}.
\label{x123int34}
 \eeqa
 In section 2, one encounters the following integral
\beqa
 \int d^2 \!z |1-z|^{a} |z|^{b} (z - \bar{z})^{c}
(z + \bar{z})^{d},
 \eeqa
 where $d=0,1,2$. The region of integration is the upper half
of the complex plane. For $d=0,1$ the integral was evaluated  in \cite{Fotopoulos:2001pt}
and for $d=2$ in \cite{Hatefi:2010ik}.
 The Mandelstam variables are defined by
\beqar
s&=&-\frac{\alpha'}{2}(k_1+k_3)^2,\qquad t=-\frac{\alpha'}{2}(k_1+k_2)^2,\qquad u=-\frac{\alpha'}{2}(k_2+k_3)^2.
\qquad\eeqar
$T_{\alpha}$ denotes the $U(N)$ generators with normalization
\beqa
\Tr(T_{\alpha}T_{\beta})=N\delta_{\alpha\beta}
\eeqa

\renewcommand{\theequation}{B.\arabic{equation}}
 \setcounter{equation}{0}
  \section*{Appendix B: Analysis of $<V_{C}V_{\phi} V_A>$
  }

In the main text, we have analyzed $<V_{C}V_{\phi} V_AV_A>$.
Here we carry out a parallel analysis for the simpler case of $<V_{C}V_{\phi} V_A>$ and we find all infinite contact terms of this amplitude.
The S-matrix element
is given by the following correlation function
\begin{eqnarray}
{\cal A}^{C\phi A} & \sim & \int dx_{1}dx_{2}dzd\bar{z}\,
  \lan V_{\phi}^{(0)}{(x_{1})}
V_{A}^{(-1)}{(x_{2})}
V_{C}^{(-\frac{1}{2},-\frac{1}{2})}(z,\bar{z})\ran,\labell{cpastring}\eeqa
Substituting the vertex operators, \reef{cpastring} can be written as
\beqa {\cal A}^{C\phi A}&\sim& \int
 dx_{1}dx_{2}dx_{4} dx_{5}\,
(P_{-}\fsH_{(n)}M_p)^{\al\be}\xi_{1i}\xi_{2a}x_{45}^{-1/4}(x_{24}x_{25})^{-1/2}\nonumber\\&&
\times(I_1+I_2)\,\labell{cpa125}\eeqa
with
\beqa
I_1&=&{<:\partial X^i(x_1)e^{\alpha'ik_1.X(x_1)}:e^{\alpha'ik_2.X(x_2)}
:e^{i\frac{\alpha'}{2}p.X(x_4)}:e^{i\frac{\alpha'}{2}p.D.X(x_5)}:>}
 \  \non \\&&\times{<:S_{\al}(x_4):S_{\be}(x_5):\psi^a(x_2):>},\nonumber\\
I_2&=&{<:e^{\alpha'ik_1.X(x_1)}:e^{\alpha'ik_2.X(x_2)}
:e^{i\frac{\alpha'}{2}p.X(x_4)}:e^{i\frac{\alpha'}{2}p.D.X(x_5)}:>}
 \  \non \\&&\times{<:S_{\al}(x_4):S_{\be}(x_5):\alpha'ik_{1b}\psi^{b}\psi^{i}(x_1):\psi^a(x_2):>}.
\label{cpp1234}
\eeqa
Using Wick's theorem, one can show that
\beqa
I_1^a&=&<:S_{\al}(x_4):S_{\be}(x_5):\psi^a(x_2):>=2^{-1/2}x_{45}^{-3/4}(x_{24}x_{25})^{-1/2}
(\gamma^{a}C^{-1})_{\alpha\beta}.\label{1opa2}
\eeqa
 The Wick-like rule \cite{Liu:2001qa,Kostelecky:1986xg} has been generalized to find the correlation function of two spin operators and some number of currents \cite{Hatefi:2010ik,Hatefi:2012wj}.

 Using these results, one can obtain the correlation function between two spin operators, one current and one world-sheet fermion as follows
\beqa
I_2^{aib}&=&<:S_{\al}(x_4):S_{\be}(x_5):\psi^b\psi^i(x_1)::\psi^a(x_2)>\nonumber\\
&=&\bigg\{(\Gamma^{aib}C^{-1})_{\alpha\beta}
+\frac{\alpha' Re[x_{14}x_{25}]}{x_{12}x_{45}}\bigg(\eta^{ab}(\gamma^{i}C^{-1})_{\alpha\beta}\bigg)\bigg\}
\nonumber\\&&\times2^{-3/2}x_{45}^{1/4}(x_{14}x_{15})^{-1}(x_{24}x_{25})^{-1/2}.
\label{6cpp8}\eeqa
Replacing the above spin correlators in the amplitude and performing the correlators over $X$, one finds:
\beqa
{\cal A}^{C\phi A}&\!\!\!\!\sim\!\!\!\!\!&\int dx_{1}dx_{2}dx_{4}dx_{5}(P_{-}\fsH_{(n)}M_p)^{\al\be}I\xi_{1i}\xi_{2a}x_{45}^{-1/4}
(x_{24}x_{25})^{-1/2}\nonumber\\&&\times\bigg(I_1^a(a^i_1)+i\alpha' k_{1b}I_2^{aib}\bigg)\labell{amp3cpa},\eeqa
where
\beqa
I&=&|x_{12}|^{ \alpha'^2k_1.k_2}|x_{14}x_{15}|^{\frac{\alpha'^2}{2} k_1.p}|x_{24}x_{25}|^{\frac{\alpha'^2}{2} k_2.p}
|x_{45}|^{\frac{\alpha'^2}{4}p.D.p},\nonumber\\
a^i_1&=&ip^{i}\frac{x_{54}}{x_{14}x_{15}}.
\label{icpa}
\eeqa
The amplitude has $SL(2,R)$ invariance; let us gauge fix it as
\beqa
(x_1,x_2,x_4,x_5)&=&(x,-x,i,-i) ,dx_{1}dx_{2}dx_{4}dx_{5}=-2i(1+x^2)dx
\eeqa
With this gauge fixing, the amplitude takes
\beqa {\cal A}^{C\phi A}&=&\int_{-\infty}^{\infty} dx (x^2+1)^{2t-1} (2x)^{-2t} (2\xi_{1i} \xi_{2a} 2^{-1/2})\nonumber\\&&\times
\bigg[-p^i\Tr(P_{-}\fsH_{(n)}M_p\gamma^{a})
+k_{1b}\Tr(P_{-}\fsH_{(n)}M_p\Gamma^{bai})\bigg]
\labell{11lcpa}\eeqa
Note that the term $\frac{\alpha' Re[x_{14}x_{25}]}{x_{12}x_{45}}$ does not have any contribution to the amplitude because the integrand is odd
but we have to take integration over the whole space time.

Having found the integral, the final result for the amplitude is given by
\beqa {\cal A}^{C\phi A}&=& (2\xi_{1i} \xi_{2a} 2^{-1/2}\pi^{1/2})\frac{\Gamma(-t+\frac{1}{2})}{\Gamma(-t+1)}\nonumber\\&&\times
\bigg[-p^i\Tr(P_{-}\fsH_{(n)}M_p\gamma^{a})
+k_{1b}\Tr(P_{-}\fsH_{(n)}M_p\Gamma^{bai})\bigg]
\labell{11lcpa}\eeqa
where $t=-(k_1+k_2)^2$ and we used the momentum conservation as well. The expansion is low energy expansion which can be
achieved by sending the Mandelstam variable $t$ to zero which means that we took  the $\al'\rightarrow 0$ limit of the string amplitude.
 One can then understand that the amplitude is non zero for $n=p$.
Thus it is clear from the  gamma function  that  the scattering amplitude has just infinite contact interactions.
The expansion of the over all factor for the above amplitude is as follows
\beqa
\sqrt{\pi}\frac{\Gamma(-t+\frac{1}{2})}{\Gamma(-t+1)}&=&\pi \sum_{n=-1}^{\infty} c_n(t)^{n+1}\eeqa
where some coefficients may be written down as
\beqa
c_{-1}=1,\quad c_{0}=2ln(2),\quad c_{1}=\frac{\pi^2}{6}+2ln(2)^2
\eeqa
To begin we focus on the Chern-Simons action (see eq. (3) of \cite{Hatefi:2012wj}).
The  interactions  will include a bulk RR field ($p$-1)-form potential , one scalar field and one gauge field in the world-volume of brane.
Let us extract the first term in  \reef{11lcpa}
\beqa {\cal A}_{1}^{C\phi A}&=& (2^{-1/2}\pi\mu_{p})(2\xi_{1i} \xi_{2a} 2^{-1/2}\pi) \sum_{n=-1}^{\infty} c_n(t)^{n+1}\nonumber\\&&\times
\bigg(\frac{32}{2p!}p^i\bigg)(\veps^v)^{a_0\cdots a_{p-1}a}H^{(p)}_{a_0\cdots a_{p-1}}
\labell{11lcpa2}\eeqa
where $(2^{-1/2}\pi\mu_{p})$ is the normalization factor. The leading contact interactions can be produced by the following
coupling
\beqa
S^{(3)}
&=& \frac{\lambda^2\mu_p}{(p)!}\int d^{p+1}\sigma (\veps^v)^{a_0\cdots a_{p}}
 p\Tr(F_{a_{0}a_{1}}\phi^i)
\prt_{i} C^{(p-1)}_{a_2\cdots a_{p}}(\sigma)
\nonumber\eeqa
The scalar comes from the Taylor expansion and field strength comes from Chern-Simons action (for more details see introduction and section 5 of \cite{Hatefi:2012wj}). Note with an integration by parts,
one can re express the above coupling as
\beqa
S^{(3)}
&=& \frac{\lambda^2\mu_p}{(p)!}\int d^{p+1}\sigma (\veps^v)^{a_0\cdots a_{p}}
\Tr(A_{a_{0}}\phi^i)
p^i  H^{(p)}_{a_1\cdots a_{p}}(\sigma)
\nonumber\eeqa
Having normalized the amplitude,  we are able to produce all infinite contact interactions for the first term of
this amplitude \reef{11lcpa} by the following field theory vertices
\beqa
S^{(3)}
&=& \frac{\lambda^2\mu_p}{(p)!}\int d^{p+1}\sigma (\veps^v)^{a_0\cdots a_{p}}
\sum_{n=0}^{\infty}c_{n}(\alpha')^{n+1}\Tr(\prt^{a_{1}}...\prt^{a_{n+1}} A_{a_{0}}\prt_{a_{1}}...\prt_{a_{n+1}}\phi^i)
p^i  H^{(p)}_{a_1\cdots a_{p}}(\sigma)
\nonumber\eeqa
where $H^{(p)}=dC^{(p-1)}$. To confirm all infinite contact interactions of the second term of \reef{11lcpa}
 we have to take into account the following contact interactions  in the low energy effective action. Indeed
the leading low energy terms of the amplitude will be
reproduced by
\beqa
S^{(4)}
&=& \frac{\lambda^2\mu_p} {(p)!}\int d^{p+1}\sigma (\veps^v)^{a_0\cdots a_{p}}
\Tr(F_{a_{0}a_{1}}D_{a_{2}}\Phi^i)
C^{(p-1)}_{ia_3\cdots a_{p}}(\sigma)p(p-1)
\nonumber\eeqa
Taking integration by parts and considering the antisymmetric property of $(\veps^v)^{a_0\cdots a_{p}} $
we can show that the all infinite contact terms of the second term of \reef{11lcpa} are reproduced by the following contact interactions
\beqa
S^{(4)}
&=& \frac{\lambda^2\mu_p} {(p)!}\int d^{p+1}\sigma (\veps^v)^{a_0\cdots a_{p}} (p)H^{(p)}_{ia_2\cdots a_{p}}(\sigma)\nonumber\\&&\times
\sum_{n=-1}^{\infty}c_{n}(\alpha')^{n+1}\Tr(\prt_{a_{1}}...\prt_{a_{n+1}}A_{a_{0}}  \prt^{a_{1}}...\prt^{a_{n+1}} D_{a_{1}}\Phi^i)
\nonumber\eeqa

\vspace{.3in}  
\section*{Acknowledgment}
E.H would like to thank K.S.Narain for very valuable discussions. He would also like to thank F.Quevedo, L.Alvarez-Gaume, N.Arkani-Hamed, G.Veneziano, I.Antoniadis and N.Lambert  for very helpful conversations.


\end{document}